\documentclass[conference]{IEEEtran}
\usepackage{amsmath,amsfonts}
\usepackage{algorithmic}
\usepackage{graphicx}
\usepackage{svg} 
\usepackage{cite}
\usepackage{array}
\usepackage[caption=false,font=normalsize,labelfont=sf,textfont=sf]{subfig}
\usepackage{textcomp}
\usepackage{booktabs}
\usepackage{stfloats}
\usepackage{pgfplots}
\usepackage{url}
\usepackage{verbatim}
\usepackage{graphicx}
\def\BibTeX{{\rm B\kern-.05em{\sc i\kern-.025em b}\kern-.08em
    T\kern-.1667em\lower.7ex\hbox{E}\kern-.125emX}}
\usepackage{balance}
\usepackage{tikz}
\usetikzlibrary{shapes}
\usepackage{pgfplots}
\usetikzlibrary{spy}
\usepackage{soul}
\usepackage{amssymb}
\usepackage[colorlinks=true, linkcolor=blue, urlcolor=blue, citecolor=blue]{hyperref}

\usepackage{pgfplots}
\pgfplotsset{compat=1.18}

\usepackage{tikz}

\usepackage{tikz}
\usepackage{multirow}
\usepackage{pgfplots} 
\usetikzlibrary{shapes}
\usetikzlibrary{plotmarks}
\usepackage{pgfplots}
\pgfplotsset{compat=1.18}
\usepackage{float}

\begin{document}
	\tikzset{new spy style/.style={spy scope={%
				magnification=2.8,
				size=1.25cm,
				connect spies,
				every spy on node/.style={
					rectangle,
					draw,
				},
				every spy in node/.style={
					draw,
					rectangle,
				}
			}
		}
	}

\title{Performance Analysis of AFDM Transceiver Under Practical IQ Imbalance}


\author{Rishikesh Bharti, Tonmoy Rajkhowa, Sanjeev Sharma, Kuntal Deka, Zilong Liu}

\maketitle
\begin{abstract}

This paper investigates the impact of in-phase/quadrature (IQ) imbalance on signal detection in affine frequency-division multiplexing (AFDM) systems over doubly dispersive Rayleigh fading channels. The bit error rate (BER) performance of AFDM under various IQ mismatch conditions is evaluated and compared with that of an ideal transceiver. Simulation results show that, under ideal conditions, BER decreases rapidly with increasing signal-to-noise ratio (SNR). However, IQ imbalance introduces image interference that spreads across the affine domain, causing significant performance degradation. 
The results further highlight the impact of practical IQ imbalance on AFDM systems across different quadrature amplitude modulation (QAM) orders, channel scenarios, and comparisons with orthogonal frequency-division multiplexing (OFDM) and orthogonal time frequency space (OTFS) systems. In particular, higher-order QAM performance degrades significantly, indicating the need for IQ imbalance compensation techniques or impairment-aware detection algorithms to ensure robust operation in practical AFDM systems.


\end{abstract}


\IEEEpeerreviewmaketitle

\section{Introduction}

Beyond-5G wireless networks are expected to operate in high-mobility environments characterized by significant delay and Doppler spreads. In such doubly dispersive channels, conventional multi-carrier waveforms such as orthogonal frequency-division multiplexing (OFDM) experience severe performance degradation due to the loss of sub-carrier orthogonality caused by Doppler effects~\cite{bemani2023affine, 11245520}. To address this limitation, affine frequency-division multiplexing (AFDM) has recently emerged as a promising waveform capable of providing reliable communication in highly dynamic time-varying channels~\cite{bemani2021afdm}.


AFDM is a chirp-based multi-carrier waveform built upon the discrete affine Fourier transform (DAFT), which generalizes the discrete Fourier transform (DFT). The DAFT is characterized by two parameters, namely the pre-chirp and post-chirp factors \cite{bemani2023affine}, which can be optimized based on the delay and Doppler characteristics of the channel. By appropriately selecting these parameters, AFDM maps data symbols onto orthogonal chirp sub-carriers, producing a waveform whose instantaneous frequency evolves according to an affine transformation. 

AFDM achieves full DD diversity in time-varying channels, enabling reliable communication in scenarios where OFDM experiences significant performance degradation due to Doppler-induced inter-carrier interference (ICI) \cite{11245520, arous2026adaptive}. Moreover, AFDM can be implemented with minimal modifications to existing OFDM transceivers by introducing a simple phase-correction block and replacing the cyclic prefix (CP) with a chirp-periodic prefix (CPP) \cite{bemani2023affine}. As a result, AFDM preserves compatibility with conventional multi-carrier communication architectures.

Despite its excellent resilience to DD dispersion, AFDM is vulnerable to mismatches in the IQ components between the transmitted and received symbols. This mismatch can destroy the sub-carrier orthogonality and introduce additional interferences that significantly impact the system performance. Therefore, analyzing the impact of IQ imbalances on AFDM systems is crucial for assessing their practical feasibility and deployment in future high-mobility 6G communication system.


In practical AFDM transceivers, IQ imbalance can significantly degrade symbol detection performance. Gain mismatches ($g_t$, $g_r$) and phase mismatches ($\phi_t$, $\phi_r$) introduce amplitude and phase distortions between the in-phase and quadrature signal components of the transmitted and received signals~\cite{11245520}. These impairments distort the chirp-domain signal structure generated by the inverse DAFT (IDAFT) at the transmitter and processed by the DAFT at the receiver, thereby reducing the effectiveness of channel equalization and symbol recovery~\cite{wu2024afdm, anoop2026multi}. Consequently, IQ imbalance introduces additional interference and leads to noticeable performance degradation compared with the ideal AFDM system.

\begin{figure*}[!htbp]
	\centering
	\includegraphics[
	width=\linewidth,
	height=15cm,
	keepaspectratio
	]{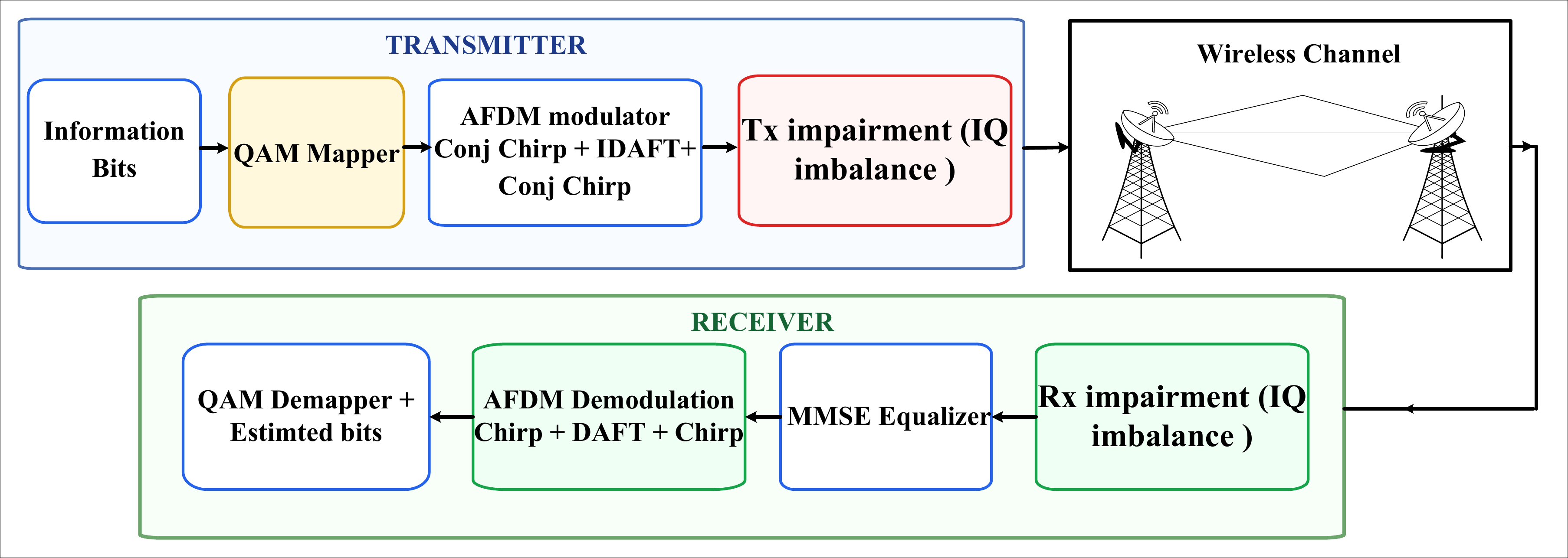}
	\caption{AFDM transceiver with IQ imbalance.}
	\label{fig:afdm_block}
\end{figure*}

To summarize, the contributions of this paper are as follows:
\begin{itemize}

\item This is the first paper that investigates the impact of IQ imbalance for signal detection task in AFDM systems and evaluate the system robustness.

\item The performance of AFDM in the presence of IQ imbalance is investigated by addressing a practical impairment that has not been comprehensively studied in the existing AFDM literature. Building upon prior studies, this work systematically analyzes the impact of IQ imbalance under realistic operating conditions and provides insights into its effect on AFDM system performance, as illustrated in Fig. \ref{illust}.

\item The bit error rate (BER) performance of the AFDM system is analyzed under IQ imbalance and compared with orthogonal time frequency space (OTFS) and OFDM systems by considering different modulation orders and wireless channel scenarios. Further, IQ imbalance compensation in the AFDM system is also considered in this paper.


\end{itemize}

    
 

This paper is organized as follows: Section~\ref{sec:preliminaries} introduces the DAFT. Section~\ref{sec:system-model} presents the AFDM system model under IQ imbalance and discusses the AFDM signal detection and Section~\ref{sec:results-discussions} analyzes the BER performance of the AFDM transceiver under IQ imbalance. Finally, Section~\ref{sec:conclusion} concludes the paper and outlines potential directions for future research.

 \section{ DAFT Preliminaries} \label{sec:preliminaries}


This section introduces the DAFT, the fundamental signal transform underlying AFDM systems. AFDM is a chirp-based multi-carrier modulation scheme specifically designed for high-mobility, doubly dispersive channels, where communication is performed in the DAFT domain \cite{bemani2021afdm}. At the transmitter, information symbols are mapped to the affine domain and then transformed to time domain through IDAFT. The signal is then transmitted over the channel and then received at the receiver where DAFT is employed to recover the transmitted symbols \cite{anoop2026multi}. Similar to the IFFT/FFT operations used in OFDM, the IDAFT/DAFT pair enables modulation and demodulation. However, the incorporation of chirp-domain processing provides enhanced resilience against delay and Doppler impairments.

A key advantage of AFDM is its low-complexity implementation. The DAFT/IDAFT operations can be efficiently realized using the conventional FFT/IFFT framework supplemented by simple phase-correction (chirp) blocks \cite{savaux2024special}. This allows AFDM to leverage existing OFDM hardware architectures while benefiting from the robustness of chirp-based signaling. To combat inter-symbol interference in doubly dispersive channels, the IDAFT-generated signal is extended with a chirp-periodic prefix (CPP), which replaces the cyclic prefix (CP) employed in OFDM \cite{li2025affine,bemani2021affine}. At the receiver, the CPP is removed before applying the DAFT, thereby transforming the received signal back into the DAFT domain for subsequent equalization and symbol detection.


Mathematically, the DAFT matrix is defined as \cite{bemani2023affine}:
\begin{equation} \small
\mathbf{A}
=
\mathbf{\Lambda}(c_{2})
\mathbf{F}
\mathbf{\Lambda}(c_{1}),
\label{eq:daft_matrix}
\end{equation}
where \(\mathbf{A}\in\mathbb{C}^{N\times N}\) denotes the DAFT matrix. In (\ref{eq:daft_matrix}), \(\mathbf{\Lambda}(c_{1})\) and \(\mathbf{\Lambda}(c_{2})\) denote diagonal chirp matrices, and \(\mathbf{F}\in\mathbb{C}^{N\times N}\) denotes the DFT matrix, whose \((m,n)\)-th element is given by:
 \begin{equation} \small 
F_{m,n} =
 \frac{1}{\sqrt{N}}
 e^{-j\frac{2\pi mn}{N}},
 \qquad
 m,n=0,1,\ldots,N-1.
 \end{equation}
  the DFT matrices size is $\mathbf{F}\in\mathbb{C}^{N\times N}$.
 
The chirp matrices are defined as
\begin{equation} \small 
\mathbf{\Lambda}(c_i)
=
\operatorname{diag}
\!\left(
e^{-j2\pi c_i n^2}
\right),
\qquad
n=0,1,\ldots,N-1,
\label{eq:chirp_matrix}
\end{equation}
where \(c_i\) denotes the chirp parameter (\(i\in\{1,2\}\)).
Consequently, $\mathbf{\Lambda}(c_i)\in\mathbb{C}^{N\times N}$.

The chirp parameters employed in this work are also illustrated in the AFDM transceiver block diagram shown in Fig.~\ref{fig:afdm_block}. The transmitter performs conjugate chirp multiplication with the IDAFT, whereas the receiver performs chirp multiplication with the DAFT.

\section{AFDM System Model Under IQ Imbalance} \label{sec:system-model}
In this section, the transmission and reception procedures of the DAFT-based AFDM waveform is detailed. 

\subsection{AFDM transmitter Under IQ Imbalance}
 The information bits, mapped to the quadrature amplitude modulation (QAM) mapper, is fed to the AFDM system that places these symbols to the chirp domain sub-carrier using the IDAFT to convert the affine-domain symbol into the time domain. Let $\mathbf{x}$ be a vector representing the QAM symbols.
\par
The transmitted signal by use of IDAFT can be defined as:
\begin{equation} \small 
\mathbf{s} = \mathbf{\Lambda}^{H}(c_{2}) \mathbf{F}^{H} \mathbf{\Lambda}^{H}(c_{1}){\mathbf{x} }
\end{equation} 
where $\mathbf{\ s }\in\mathbb{C}^{N\times 1}$ is the transmitted signal vector in the time domain.  $c_1$ and $c_2$ represent the post-chirp and pre-chirp parameters, respectively. These parameters are tunable chirp coefficients that determine the time-frequency characteristics of the AFDM waveform. To mitigate inter-symbol interference (ISI), AFDM employs a CPP, which serves a role similar to the CP used in OFDM systems while preserving the chirp structure of the transmitted signal \cite{li2023review}.

In practical scenarios, the transmitted signal always experience IQ imbalances which can be modeled as\cite{singh2023deep}:
\begin{equation} \small 
\tilde{\mathbf{s}}
=
\alpha_{t1}\mathbf{s}
+
\alpha_{t2}\mathbf{s}^{*},
\label{eq:tx_impairment}
\end{equation}
where $\alpha_{t1}$ and $\alpha_{t2}$ represents the transmitter IQ imbalance coefficients which can be further defined as:
\begin{align} \label{coe} \small
\alpha_{t1}
&=
\cos\left(\frac{\phi_t}{2}\right)
+
jg_t\sin\left(\frac{\phi_t}{2}\right) \nonumber \\
\alpha_{t2}
&=
g_t\cos\left(\frac{\phi_t}{2}\right)
-
j\sin\left(\frac{\phi_t}{2}\right),
\end{align}
where $g_t$ and $\phi_t$ denote the gain and phase mismatch parameters, respectively.

\subsection{Doubly Selective Channel}
The AFDM is designed for linear time-varying (LTV) multi-path channels with multiple DD components \cite{rou2024orthogonal}. This channel is expressed as:
\begin{equation} \small 
\mathbf{H}
=
\sum_{l=1}^{L}
h_l\mathbf{\Gamma}_{\mathrm{CPP}_l}
\mathbf{\Delta}(\nu_l)
\mathbf{\Pi}(\tau_l) \quad \in\mathbb{C}^{N\times N}
\label{eq:channel_model}
\end{equation}
where $h_l$ denotes the complex gain of the $l$-th propagation path.  $\tau_l$ and $\nu_l$ represent the corresponding delay and Doppler shift, respectively.
The parameter $L$ represents the total number of propagation paths between the transmitter (Tx) and receiver (Rx). The circular delay-shift marix ($\mathbf{\Pi}(\tau_l) \in \mathbb{R}^{N \times N}$),  the CPP matrix $\mathbf{\Gamma}_{\mathrm{CPP}_l}\in \mathbb{C}^{N \times N}$ and the Doppler phase-rotation matrix ($\mathbf{\Delta}(\nu_l) \in \mathbb{C}^{N \times N}$), can be written as:

\begin{equation} \small
  \mathbf{\Pi}(\tau_l) = \begin{bmatrix} 
0 & \cdots & 0 & 1 \\ 
1 & \cdots & 0 & 0 \\ 
\vdots & \ddots & \ddots & \vdots \\ 
0 & \cdots & 1 & 0 
\end{bmatrix}_{N \times N}  ,
\end{equation}

\begin{equation}
\small
\mathbf{\Gamma}_{\mathrm{CPP}_l}
=
\operatorname{diag}
\left(
\begin{cases}
e^{-j2\pi c_1\left(N^2-2N(l-n)\right)}, & n<l,\\[2mm]
1, & n\ge l,
\end{cases}
\right), \text{and}
\end{equation}
\begin{equation} \small 
\mathbf{\Delta}(\nu_l) \triangleq 
\operatorname{diag}
\left(
e^{-j2\pi \nu_ln}
\right),
\quad n = 0,1,\ldots,N-1.
\end{equation}




\subsection{Receiver Hardware Impairments}
The received signal, along with the channel effects, can be expressed as:
\begin{equation} \small 
\mathbf{r}
=
\mathbf{H}
\tilde{\mathbf{s}}
+
\mathbf{w}
\label{eq:rx_signal},
\end{equation}


\noindent where $\mathbf{\ r}\in\mathbb{C}^{N\times 1}$ is the received signal and $\mathbf{w}\sim\mathcal{CN}(0,\sigma_n^2\mathbf{I})$ denotes complex additive white gaussian noise (AWGN) with zero mean and $\sigma_n^2$ variance.  $\mathbf{I}$ is the identity matrix of size $N \times N$. The Rx removes the CPP and applies the DAFT to map the receive signal back into the affine domain. This signal $\mathbf{r}$ is further altered as $\tilde{\mathbf{r}}$ by the IQ imbalances at the Rx, which can be modeled as~\eqref{coe}:

\begin{equation} \small 
\tilde{\mathbf{r}}
=
\alpha_{r1}\mathbf{r}
+
\alpha_{r2}\mathbf{r}^{*},
\end{equation}
where $\alpha_{r1}$ and $\alpha_{r2}$ denote the Rx IQ imbalance coefficients and are defined similarly to the transmitter IQ imbalance coefficients.


\subsection{Signal Representation}
The AFDM signal is represented by the affine-domain vector $\mathbf{x}$, which is transformed into the time-domain signal $\mathbf{s}$ for transmission through a doubly dispersive channel in the presence of AWGN. Due to transceiver IQ imbalance, an additional image component is received at the Rx. The received signal $\tilde{\mathbf{r}}$ is processed by the MMSE-based detector to detect our signal $\mathbf{x}$.

\subsection{MMSE-based Detector}



Following DAFT-domain reception, signal detection is carried out using an minimum mean square detector (MMSE) equalizer to estimate the transmitted symbols \cite{li2024chirp}. The MMSE equalizer mitigates both channel distortion and additive noise by minimizing the mean square error between the transmitted and detected symbols. The corresponding MMSE equalization matrix is expressed as:

 \begin{equation} \small 
\mathbf{W}_{\rm{MMSE}}
=
\left(
\mathbf{H}^{H}\mathbf{H}
+
\sigma_n^2\mathbf{I}
\right)^{-1}
\mathbf{H}^{H}
\label{eq:mmse},
\end{equation}

\noindent where $\mathbf{W}_{\rm MMSE}\in\mathbb{C}^{N\times N}$ denotes the MMSE equalization matrix, $\mathbf{H}$ represents the effective channel matrix in the DAFT domain.

The received signal is then equalized as:

\begin{equation} \small 
\mathbf{z}
=
\mathbf{W}_{\rm{MMSE}}
\tilde{\mathbf{r}},
\end{equation}

\noindent where $\tilde{\mathbf{r}}\in\mathbb{C}^{N\times 1}$ is the received DAFT-domain signal and $\mathbf{z}\in\mathbb{C}^{N\times 1}$ denotes the equalized signal vector.

After equalization, AFDM demodulation is performed by applying the inverse DAFT-domain processing, yielding the estimated symbol vector:

\begin{equation} \small 
\hat{\mathbf{s}}
=
\left(
\mathbf{\Lambda}_{c_2}
\text{F}
\mathbf{\Lambda}_{c_1}
\mathbf{z}
\right)
\label{eq:afdm_rx}
\end{equation}

\noindent where $\hat{\mathbf{s}}\in\mathbb{C}^{N\times 1}$ contains the detected data symbols. Finally, the estimated symbols are demapped to recover the transmitted information bits and compute the BER performance of the AFDM system.


\section{Results and Discussion} \label{sec:results-discussions}


This section evaluates the BER performance of AFDM in both ideal and IQ-impaired scenarios and provides a comparative analysis with conventional OFDM and OTFS systems. Performance is assessed over Rayleigh fading and delay-Doppler channels using MMSE-based signal detection. The considered cases  and simulation parameters  are summarized in TABLE~\ref{tab:cases_separated} and \ref{tab:sim_parameters}.

\begin{table}[htb]
\centering
\caption{Simulation Cases for Transceiver Impairments}
\label{tab:cases_separated}
\scalebox{0.9}{
\begin{tabular}{lccccccc}
\toprule
\textbf{Case} & \boldmath{$g_t$} & \boldmath{$g_r$} & \boldmath{$\phi_t$} & \boldmath{$\phi_r$}   & \textbf{Remarks} \\
\midrule
Ideal & 1 & 1 & $0^\circ$ & $0^\circ$  & No impairments \\
Case 1 & 0.2 & 0 & $5^\circ$ & $0^\circ$ &  Transmitter I/Q imbalance\\
Case 2 & 0.2 & 0.2 & $5^\circ$  & $5^\circ$   &   Both transmitter and receiver I/Q imbalance     \\
\bottomrule
\end{tabular}}
\end{table}

\subsection{BER Performance in Rayleigh Fading Channel}

Fig.~\ref{3-tap Rayleigh} compares the BER performance of AFDM, OTFS, and OFDM under transmitter IQ imbalance with gain mismatch $g_t=0.2$ and phase mismatch $\phi_t=5^\circ$ over a 3-tap Rayleigh fading channel. The analysis considers both 4-QAM and 16-QAM modulation schemes, with zero delay and Doppler shifts. For AFDM and OFDM, a block size of $N=64$ and CP length of $8$ are used, whereas OTFS employs $16$ delay bins and $16$ Doppler bins. As expected, the BER of all waveforms decreases with increasing SNR due to the reduced impact of additive noise.

For 4-QAM modulation, AFDM and OTFS outperform OFDM by exploiting greater diversity gains. In AFDM, the chirp-based DAFT processing spreads information symbols across the transmission block, while OTFS leverages delay-Doppler domain modulation to effectively utilize multipath diversity. Consequently, both waveforms achieve superior reliability compared to OFDM. The performance of AFDM is comparable to that of OTFS and is influenced by the selection of the DAFT parameters $c_1$ and $c_2$.

\begin{figure}[!hbt]
	\centering
	\includegraphics[width=.9\linewidth]{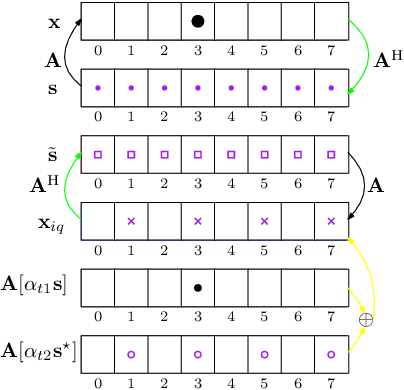}
	\caption{Impact of IQ imbalance illustration in time and affine  domains in AFDM system.}
	\label{illust}
\end{figure}

In contrast, for 16-QAM modulation, OFDM achieves lower BER than both AFDM and OTFS in the presence of IQ imbalance. This behavior arises because IQ-induced image interference spreads more extensively in the affine and delay-Doppler domains than in the conventional frequency domain. As a result, the impact of IQ imbalance outweighs the diversity advantage of AFDM and OTFS for higher-order modulation. Moreover, AFDM generates stronger image components, as illustrated in Fig.~\ref{illust}, leading to a more pronounced performance degradation.

To mitigate this impairment, a widely linear minimum mean square error (WLMMSE) compensator is applied to the AFDM 4-QAM Case-1 scenario. As shown by the green curve in Fig.~\ref{3-tap Rayleigh}, the WLMMSE compensator substantially improves the BER performance, enabling AFDM to achieve performance close to that of the ideal IQ-balanced system. \cite{tandur2010efficient}

			\pgfplotsset{every semilogy axis/.append style={
					line width=0.7 pt, tick style={line width=0.7pt}}, width=9cm,height=8cm, 
				legend style={font=\scriptsize},
				legend pos= south west}
			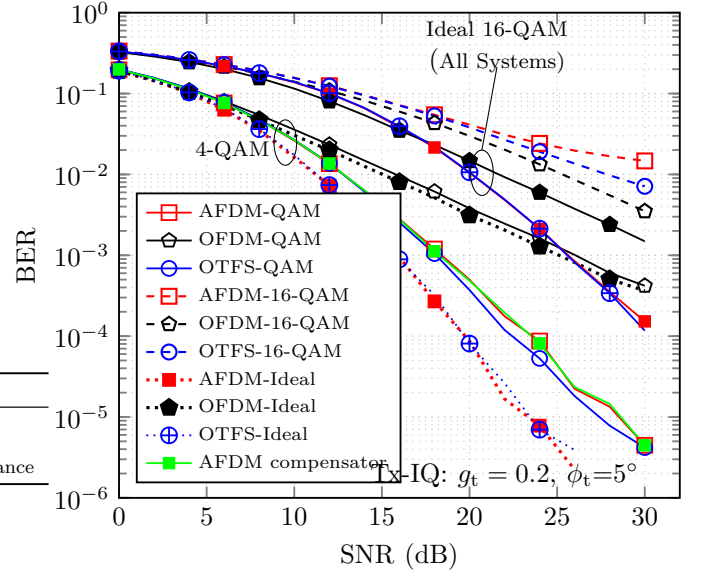
\begin{figure}[!htb]
				\centering
				\begin{tikzpicture}[new spy style]
					\begin{semilogyaxis}[xmin=0, xmax=32, ymin=1e-6, ymax=1,
						xlabel={ SNR (dB)},
						ylabel={BER},
						grid=both,
						grid style={dotted},
						legend cell align=left,
						legend entries={{AFDM-QAM}, {OFDM-QAM}, {OTFS-QAM}, {AFDM-16-QAM}, {OFDM-16-QAM}, {OTFS-16-QAM}, {AFDM-Ideal}, {OFDM-Ideal}, {OTFS-Ideal},{AFDM compensator}},
						cycle list name=black white
						]
						\coordinate (spypoint) at (axis cs:7,-700);
						\coordinate (spyviewer) at (axis cs:3.7,-1);
						
						\addplot [red, mark=square, mark size=2.8 pt, mark repeat=3]table [x={snr},y={afdm_4}]
						{results_ray3taps.txt};
						\addplot [black, mark=pentagon, mark repeat=3,mark size=2.8 pt]table [x={snr},y={ofdm_4}]
						{results_ray3taps.txt};
						\addplot [blue, mark=o, mark repeat=3,mark size=2.8 pt]table [x={snr},y={otfs_4}]{results_ray3taps.txt};
						\addplot [red, thick,  mark=square,dashed, mark repeat=3, mark size=2.8pt, mark options={scale=1,solid}]table [x={snr},y={afdm_16}]
						{results_ray3taps.txt};		
                        \addplot [black,  thick, mark=pentagon,dashed,mark repeat=3, mark size=2.8 pt, mark options={scale=1,solid}]table [x={snr},y={ofdm_16}]
						{results_ray3taps.txt};	
                           \addplot [blue,  thick, mark=o,dashed, mark repeat=3, mark size=2.8 pt, mark options={scale=1,solid}]table [x={snr},y={otfs_16}]
						{results_ray3taps.txt};	

\addplot [red, mark=square*, very thick, dotted,  mark size=2 pt, mark repeat=3, mark options={scale=1,solid}]table [x={snr},y={afdm_4}]{results_ideal_3tap_Ray_4qam.txt};		
\addplot [black, mark=pentagon*, very thick, dotted, mark repeat=2, mark size=3 pt, mark options={scale=1,solid}]table [x={snr},y={ofdm_4}]{results_ideal_3tap_Ray_4qam.txt};	
\addplot [blue, mark=oplus, mark repeat=2, dotted, mark size=3 pt, mark options={scale=1,solid}]table [x={snr},y={otfs_4}]{results_ideal_3tap_Ray_4qam.txt};	
\addplot [green, mark=square*, mark size=2 pt, mark repeat=3, mark options={scale=1,solid}]table [x={snr},y={value}]{NEW_VALUES.tex};
\addplot [red, mark=square*, mark size=2 pt, mark repeat=3, mark options={scale=1,solid}]table [x={snr},y={afdm_16}]{results_ideal_3tap_Ray.txt};	

\addplot [black, mark=pentagon*, mark repeat=2, mark size=3 pt, mark options={scale=1,solid}]table [x={snr},y={ofdm_16}]{results_ideal_3tap_Ray.txt};	
\addplot [blue, mark=oplus, mark repeat=2, mark size=3 pt, mark options={scale=1,solid}]table [x={snr},y={otfs_16}]{results_ideal_3tap_Ray.txt};	
					\end{semilogyaxis}	
					\node at (5.1,.3) {Tx-IQ: {$g_{\rm t}=0.2$,  $\phi_{\rm t}$=$5^\circ$}};
   \draw[] (4.8,4.3)  ellipse (.15 and 0.3);
    \draw[] (2.2,4.7)node[ align=center,xshift=-7mm, yshift=-1mm]{\footnotesize{4-QAM}}  ellipse (.15 and 0.3);
   
    \draw[black] (5,5.7)  node[ align=center, yshift=2.5mm]{\footnotesize{Ideal 16-QAM}\\ (\footnotesize{All Systems})}--(4.8,4.6);
				\end{tikzpicture}
				\caption{ \footnotesize{BER versus SNR comparison of AFDM, OTFS, and OFDM under transmitter IQ imbalance over a 3-tap Rayleigh fading channel with $4$-QAM and $16$-QAM.}}
				\label{3-tap Rayleigh}
			\end{figure}


	\pgfplotsset{every semilogy axis/.append style={
		line width=0.7 pt, tick style={line width=0.7pt}}, width=9cm,height=8cm, 
	legend style={font=\scriptsize},
	legend pos= south west}
\begin{figure}[!htb]
	\centering
	\begin{tikzpicture}[new spy style]
		\begin{semilogyaxis}[xmin=8, xmax=30, ymin=5e-6, ymax=.5,
			xlabel={ SNR (dB)},
			ylabel={BER},
			grid=both,
			grid style={dotted},
			legend cell align=left,
			legend entries={{AFDM-Case~1}, {OFDM-Case~1}, {OTFS-Case~1}, {AFDM-Ideal}, {OFDM-Ideal}, {OTFS-Ideal}, {AFDM-Case~2}, {OFDM-Case~2}, {OTFS-Case~2}},
			cycle list name=black white
			]
			\coordinate (spypoint) at (axis cs:7,-700);
			\coordinate (spyviewer) at (axis cs:3.7,-1);
			
			\addplot [red, mark=square, mark repeat=3, mark size=3 pt]table [x={snr},y={afdm_4}]
			{results_set1.txt};
			\addplot [black, mark=pentagon, mark repeat=3, mark size=3 pt]table [x={snr},y={ofdm_4}]
			{results_set1.txt};
			\addplot [blue, mark=o, mark size=3 pt, mark repeat=3]table [x={snr},y={otfs_4}]{results_set1.txt};
			\addplot [red, mark=square,dashed, mark size=2 pt, mark repeat=3, mark options={scale=1,solid}]table [x={snr},y={afdm_4}]
			{results_ideal.txt};		
			\addplot [black, mark=pentagon,dashed, mark repeat=2, mark size=3 pt, mark options={scale=1,solid}]table [x={snr},y={ofdm_4}]
			{results_ideal.txt};	
			\addplot [blue, mark=o,dashed, mark repeat=2, mark size=3 pt, mark options={scale=1,solid}]table [x={snr},y={otfs_4}]
			{results_ideal.txt};	
			
			\addplot [red, mark=square*, mark size=2 pt, mark repeat=3,  mark options={scale=1,solid}]table [x={snr},y={afdm_4}]{results_both.txt};		
			\addplot [black, mark=pentagon*, mark size=2 pt, mark repeat=3, mark options={scale=1,solid}]table [x={snr},y={ofdm_4}]
			{results_both.txt};	
			\addplot [blue, mark=oplus, mark size=2 pt, mark repeat=3, mark options={scale=1,solid}]table [x={snr},y={otfs_4}]
			{results_both.txt};	
			
		\end{semilogyaxis}	
		\draw[] (6.4,4.3) node[xshift=-10mm, yshift=8mm, align=center]{\footnotesize{Both Tx and Rx} \\ \footnotesize{$g_{\rm t}$=$g_{\rm r}$=$0.2$, $\phi_{\rm t}$=$\phi_{\rm r}$=$5^\circ$}} ellipse (.2 and 0.4);
		
		
	\end{tikzpicture}
	\caption{\footnotesize{BER versus SNR comparison of AFDM, OTFS, and OFDM  under 3-tap delay-Doppler Rayleigh fading channel with $4$-QAM.}}
	\label{impact}
\end{figure}
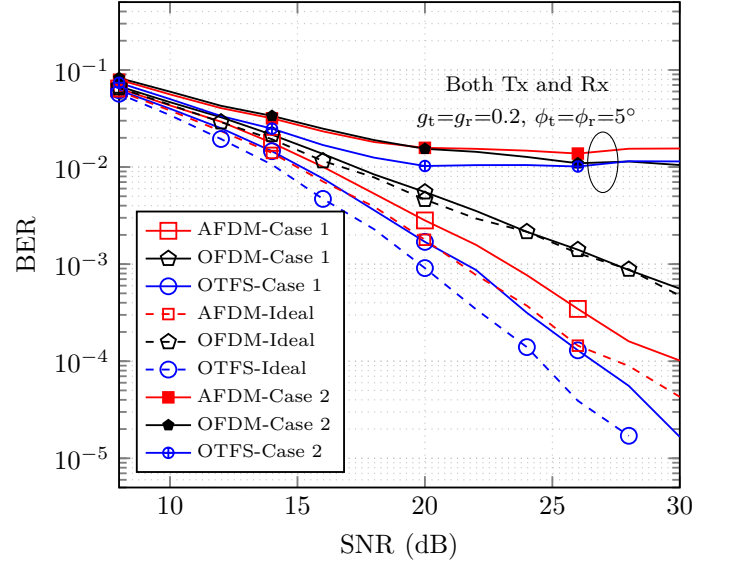

\subsection{BER Performance in Delay-Doppler Channel}

The BER performance of the proposed AFDM system is evaluated and compared with conventional OFDM and OTFS systems under a doubly selective delay-Doppler Rayleigh fading channel in the presence of IQ imbalance. 

\begin{table}[t]
\centering
\caption{Simulation Parameters}
\label{tab:sim_parameters}
\renewcommand{\arraystretch}{1.15}
\begin{tabular}{|p{3.5cm}|p{4.5cm}|}
\hline
\textbf{Parameter} & \textbf{Value} \\
\hline
Carrier frequency ($f_c$) & $6$ GHz \\
\hline
Subcarrier spacing ($\Delta f$) & $30$ kHz \\
\hline
Symbol duration ($T$) & $33.33~\mu$s \\
\hline
User velocity ($v$) & $500$ km/h \\
\hline
Maximum Doppler frequency ($f_{D,\max}$) & $2777.8$ Hz \\
\hline
Channel model & 3-path delay-Doppler Rayleigh fading \\
\hline
Path delays ($\tau$) & $[5,\;500,\;4000]$ ns \\
\hline
Discrete delay taps ($l_p$) & $[1,\;1,\;2]$ \\
\hline
Discrete Doppler taps ($k_p$) & $\left\lceil \nu_pKT \right\rceil$ \\
\hline
Doppler shift & $\nu_p=f_{D,\max}\cos(\theta_p)$ \\
\hline
Doppler angle ($\theta_p$) & Uniform over $[-\pi,\pi]$ \\
\hline
 Transmitter and receiver IQ imbalance & $g_t=g_r=0.2$, $\phi_t=\phi_r=5^\circ$ \\
\hline
AFDM block length & $N=128$ \\
\hline
AFDM chirp parameters & $c_1=c_2=\frac{1}{2N}$ \\
\hline
OTFS delay and Doppler bins & $M=16$, $N=32$ \\
\hline
OTFS frame size & $MN=512$ symbols \\
\hline
Modulation & 4-QAM, 16-QAM\\
\hline
\end{tabular}
\end{table}

Both transmitter and receiver IQ imbalance are modeled by considering an amplitude mismatch of $g_t=g_r=0.2$ and a phase mismatch of $\phi_t=\phi_r=5^\circ$.
For OFDM and AFDM transmission, a block length of $N=128$ symbols is considered. AFDM modulation employs the affine Fourier transform with chirp parameters $c_1=c_2=\frac{1}{2N}$\footnote{The AFDM chirp parameters can be optimized according to the channel characteristics}.
Similarly, for OTFS transmission, the delay and Doppler dimensions are set to $M=16$ and $N=32$, respectively, yielding a total frame size of $MN=512$ symbols. A 4-QAM modulation scheme is employed for all systems.

Fig.~\ref{impact} illustrates the BER performance comparison of AFDM, OTFS, and OFDM under an ideal transceiver configuration without IQ impairment, transmitter IQ imbalance with $g_t=0.2$ and $\phi_t=5^\circ$ (Case~1), and joint transmitter and receiver IQ imbalance with $g_t=g_r=0.2$ and $\phi_t=\phi_r=5^\circ$ (Case~2).
It can be observed that the ideal transceiver configuration provides the lowest BER for all waveforms since no additional distortion is introduced by the IQ imbalance. When only transmitter IQ imbalance is considered (Case~1), the BER performance degrades due to the image interference generated by the mismatch between the in-phase and quadrature branches. The degradation becomes more severe in the presence of both transmitter and receiver IQ imbalance (Case~2), where the combined amplitude and phase mismatches introduce additional signal distortion, resulting in an increased BER and the appearance of an error floor at high SNR regions.

Moreover, under ideal transceiver and transmitter-only IQ imbalance conditions, OFDM performance degrades more significantly than AFDM and OTFS. However, under joint transmitter and receiver IQ imbalance (Case~2), the BER performance of all systems tends to saturate because the increased image interference in AFDM and OTFS dominates their diversity gain advantage, as observed in Fig.~\ref{impact}.

\begin{figure}[!hbt]
	\centering
	\includegraphics[width=\linewidth]{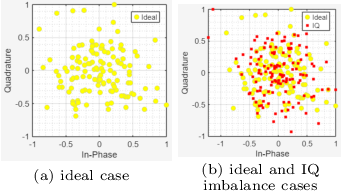}
	\caption{Impact of IQ imbalance ($g_t=1.15$ and $\phi_t=7^{\circ}$) in AFDM system.}
	\label{fig_points}
\end{figure}

\begin{figure*}[!hbt]
	\centering
	\includegraphics[width=\linewidth]{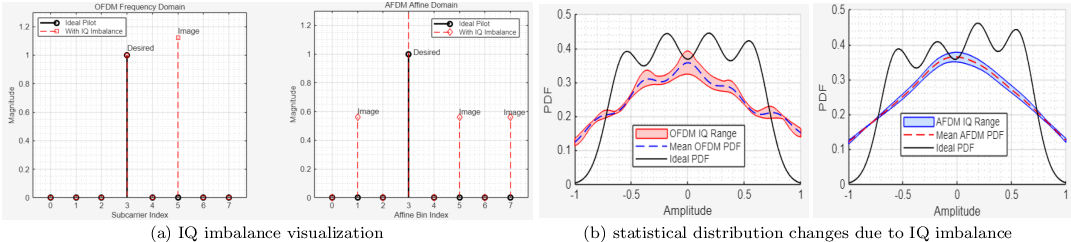}
	\caption{Visualization and statistical analysis of IQ imbalance in OFDM and AFDM systems.}
	\label{fig_pdf}
\end{figure*}
\subsection{ Impact of IQ Imbalance}

The impact of transmitter IQ imbalance on AFDM can be understood by examining the affine-domain representation of the transmitted signal. Accordingly, the IQ-impaired AFDM signal in the affine domain is expressed as:

\begin{equation} \small 
\mathbf{x}_{iq} = \mathbf{A} \tilde{\mathbf{s}}= \mathbf{A}[\alpha_{t1}\mathbf{s}+
\alpha_{t2}\mathbf{s}^{*}]=\alpha_{t1} \mathbf{A} \mathbf{s}+\alpha_{t2} \mathbf{A} \mathbf{s}^{*}
\end{equation} 

\noindent where $\mathbf{A}=\mathbf{\Lambda}(c_{2}) \mathbf{F} \mathbf{\Lambda}(c_{1})$, and 

\begin{equation} \small 
\mathbf{x}_{iq}=\alpha_{t1} \underbrace{\mathbf{A} \mathbf{A}^H}_{\mathbf{I}} \mathbf{x}+\alpha_{t2} \mathbf{A} [\mathbf{A}\mathbf{x}]^{*}=\alpha_{t1} \mathbf{x} +\alpha_{t2} \mathbf{A} \mathbf{A}^{*} \mathbf{x}^{*}.
\end{equation}

In the above expression, $\mathbf{A} \mathbf{A}^{*}=\mathbf{\Lambda}(c_{2}) \mathbf{F} \mathbf{F}^{*} \mathbf{\Lambda}^{*}(c_{2})$, which is a non-diagonal matrix. Therefore, conjugation destroys the affine orthogonality, and the transformed conjugate basis no longer aligns with individual affine bins. Consequently, IQ imbalance introduced in the time domain results in image spreading in the affine domain of AFDM systems, as observed in Fig. \ref{illust}.
For example, a single pilot symbol placed in the affine domain (at the 4th bin of $\mathbf{x}$) produces multiple symbols in the affine domain (represented by $\mathbf{x}_{iq}$) when IQ imbalance is present, as illustrated in Fig. \ref{illust}. This image spreading creates severe interference and degrades the performance of the AFDM system.





In Fig. \ref {fig_points}, the impact of IQ imbalance is analyzed using the 16-QAM constellation points for the AFDM system. The constellation points are generated in the affine domain, while IQ imbalance are introduced in the time domain. Subsequently, the time-domain signal is transformed into the affine domain using the DAFT matrix, and the in-phase and quadrature components are plotted in Fig. \ref{fig_points}.
After transforming the signal into the affine domain, IQ imbalance manifests as image interference, which degrades the orthogonality and sparsity properties of AFDM symbols. as illustrated in Fig. \ref{fig_points}. Consequently, these impairments distort the desired signal structure of the AFDM system; specifically, IQ imbalance causes energy spreading and mirror leakage. As a result, IQ impairments lead to degraded BER and sum-rate performance of AFDM system.

IQ imbalance ($g_t=1.12$ and $\phi_t=5^{\circ}$) is visualized in Fig. \ref{fig_pdf}(a). For this analysis, 8 subcarriers and affine bins are considered in the OFDM and AFDM systems, respectively, with a pilot located at index 3 and pilot value equal to one.
The results show that image components appear at the mirror subcarrier (5) in the OFDM system, whereas in the AFDM system the image energy spreads into affine odd bins\footnote{Although the energy spreads across all affine bins, a significant portion of the energy is concentrated in the odd-index bins. Furthermore, when the pilot is placed at an even index, the energy spreading predominantly appears in the even bins in the AFDM system.} (1, 5, and 7) instead of remaining localized as in OFDM. Thus, AFDM transforms the conjugate component of the desired signal into distributed affine-domain interference because the DAFT basis loses orthogonality under conjugation.
Consequently, AFDM experiences image spreading rather than localized mirror interference, as observed in the OFDM system, which requires additional attention and dedicated compensation approaches for reliable AFDM system design.

Further, in Fig. \ref{fig_pdf}(b), the statistical impact of IQ imbalance in the AFDM system is analyzed using the probability density function (PDF) of the real component of the 16-QAM AFDT signal by varying the IQ imbalance values (from ($g_t=1$ and $\phi_t=0^{\circ}$)  to $g_t=1.19$ and $\phi_t=10^{\circ}$).
The results show that IQ imbalance alters the variance and distribution shape of the AFDM signal, as observed in Fig. \ref{fig_pdf}(b). Additionally, a comparison with the OFDM signal under the same IQ imbalance conditions is also presented in Fig. \ref{fig_pdf}(b).
It can be observed that OFDM and AFDM exhibit different statistical responses to IQ imbalance. In this analysis, both OFDM and AFDM systems employ 1024 subcarriers and affine bins, respectively.


\section{Conclusion} \label{sec:conclusion}
In this paper, the impact of practical IQ imbalance on the AFDM transceiver's performance was investigated through BER-versus-SNR analysis. Impairement analysis of gain mismatch and phase imbalance was incorporated into the AFDM system model, and their effects on BER performance were evaluated. Simulation results show that the ideal AFDM system achieves excellent BER performance at high SNR. The presence of IQ imbalance significantly degrades performance and introduces an error floor in the high-SNR region also analyse the visualisation and statistical analysis of IQ imbalance in OFDM and AFDM systems. Further, IQ imbalance affects higher-order QAM more significantly than lower-order QAM in the AFDM system, as observed in the results.

\bibliographystyle{IEEEtran}
\bibliography{ieeeabr}

@article{tandur2010efficient,
  title={Efficient compensation of transmitter and receiver {IQ} imbalance in {OFDM} systems},
  author={Tandur, Deepaknath and Moonen EURASIP Member, Marc},
  journal={EURASIP J. on Advances in Signal Proc.},
  volume={2010},
  number={1},
  pages={106562},
  year={2010},
  publisher={Springer}
}

@ARTICLE{11245520,
  author={Rajkhowa, Tonmoy and Sharma, Sanjeev and Deka, Kuntal},
  journal={IEEE Wireless Commun. Lett.}, 
  title={Automatic Modulation Classification for Hardware-Impaired {OTFS} Transceiver Systems}, 
  year={2025},
  volume={},
  number={},
  pages={1-1},
  doi={10.1109/LWC.2025.3632246}}

@inproceedings{bemani2021afdm,
  title={{AFDM}: A full diversity next generation waveform for high mobility communications},
  author={Bemani, Ali and Ksairi, Nassar and Kountouris, Marios},
  booktitle={2021 IEEE Int. Conf. Commun. Workshops (ICC Workshops)},
  pages={1--6},
  year={2021},
  organization={IEEE}
}

@article{bemani2023affine,
  title={Affine frequency division multiplexing for next generation wireless communications},
  author={Bemani, Ali and Ksairi, Nassar and Kountouris, Marios},
  journal={IEEE Trans. Wireless Commun.},
  volume={22},
  number={11},
  pages={8214--8229},
  year={2023},
  publisher={IEEE}
}

@article{anoop2026multi,
  title={Multi-mode Index Modulation based on Affine Frequency Division Multiplexing},
  author={Anoop, A and Thomas, Christo Kurisummoottil and Kala, S and Benifa, JV Bibal and Saad, Walid},
  journal={Phys. Commun.},
  pages={103059},
  year={2026},
  publisher={Elsevier}
}

@article{savaux2024special,
  title={Special cases of {DFT}-based modulation and demodulation for affine frequency division multiplexing},
  author={Savaux, Vincent},
  journal={IEEE Trans. Commun.},
  volume={72},
  number={12},
  pages={7627--7638},
  year={2024},
  publisher={IEEE}
}

@article{li2025affine,
  title={Affine frequency division multiplexing over wideband doubly-dispersive channels with time-scaling effects},
  author={Li, Xiangxiang and Wang, Haiyan and Ge, Yao and Shen, Xiaohong and Guan, Yong Liang and Wen, Miaowen and Yuen, Chau},
  journal={IEEE Trans. Wireless Commun.},
  year={2025},
  publisher={IEEE}
}

@article{arous2026adaptive,
  title={Adaptive Interference Suppression for {AFDM} Channel Estimation Under OFDM Coexistence},
  author={Arous, Abdelali and Haif, Hamza and Arslan, H{\"u}seyin},
  journal={IEEE Wireless Commun. Lett.},
  volume={15},
  pages={1792--1796},
  year={2026},
  publisher={IEEE}
}

@article{li2023review,
  title={A review on orthogonal time--frequency space modulation: State-of-art, hotspots and challenges},
  author={Li, Mao and Liu, Wei and Lei, Jing},
  journal={Comput. Netw.},
  volume={224},
  pages={109597},
  year={2023},
  publisher={Elsevier}
}

@article{singh2023deep,
  title={Deep learning-assisted {OFDM} detection with hardware impairments},
  author={Singh, Amit and Sharma, Sanjeev and Deka, Kuntal and Bhatia, Vimal},
  journal={J. Commun. Inf. Netw.},
  volume={8},
  number={4},
  pages={378--388},
  year={2023},
  publisher={PTP}
}

@article{wu2024afdm,
  title={{AFDM} signal detection based on message passing scheme},
  author={Wu, Lifan and Luo, Shan and Song, Dongxiao and Yang, Fan and Lin, Rongping and Xie, Siyu},
  journal={Digit. Signal Process.},
  volume={153},
  pages={104633},
  year={2024},
  publisher={Elsevier}
}

@article{li2024chirp,
  title={Chirp parameter selection for affine frequency division multiplexing with MMSE equalization},
  author={Li, Zunqi and Zhang, Chuanbin and Song, Ge and Fang, Xiaojie and Sha, Xuejun and Slock, Dirk TM},
  journal={IEEE Trans. Commun.},
  volume={73},
  number={7},
  pages={5079--5093},
  year={2024},
  publisher={IEEE}
}

@article{rou2024orthogonal,
  title={From orthogonal time--frequency space to affine frequency-division multiplexing: A comparative study of next-generation waveforms for integrated sensing and communications in doubly dispersive channels},
  author={Rou, Hyeon Seok and De Abreu, Giuseppe Thadeu Freitas and Choi, Junil and Gonz{\'a}lez, David and Kountouris, Marios and Guan, Yong Liang and Gonsa, Osvaldo},
  journal={IEEE Signal Process. Mag.},
  volume={41},
  number={5},
  pages={71--86},
  year={2024},
  publisher={IEEE}
}

@inproceedings{bemani2021affine,
  title={Affine frequency division multiplexing for next-generation wireless networks},
  author={Bemani, Ali and Cuozzo, Giampaolo and Ksairi, Nassar and Kountouris, Marios},
  booktitle={2021 17th Int. Symp. Wireless Commun. Syst. (ISWCS)},
  pages={1--6},
  year={2021},
  organization={IEEE}
}
\end{document}